\documentclass[fleqn,twoside]{article}
\usepackage[headings]{espcrc2}

\usepackage{graphicx}
\usepackage[figuresright]{rotating}
\usepackage{pstricks}
\usepackage{amsmath}

\newcommand{\lbl}[1]{\label{eq:#1}}
\newcommand{ \rf}[1]{(\ref{eq:#1})}

\newcommand{\be}{\begin{equation}}
\newcommand{\ee}{\end{equation}}
\newcommand{\bea}{\begin{eqnarray}}
\newcommand{\eea}{\end{eqnarray}}

\newcommand{\ra}{\rightarrow}

\newcommand{\lesssim}{ {\
\lower-1.2pt\vbox{\hbox{\rlap{$<$}\lower5pt\vbox{\hbox{$\sim$}}}}\ } 
}
\newcommand{\gtrsim}{ {\
\lower-1.2pt\vbox{\hbox{\rlap{$>$}\lower5pt\vbox{\hbox{$\sim$}}}}\ } 
}

\newcommand{\cL}{{\cal L}}

\newcommand{\cO}{{\cal O}}

\newcommand{\Imm}{\mbox{\rm Im}}

\newcommand{\tr}{\mbox{\rm tr}}

\newcommand{\MeV}{\mbox{\rm MeV}}

\newcommand{\annd}{\mbox{\rm and}}

\newcommand{\bfw}{\mbox{\bf w}}
\newcommand{\bfg}{\mbox{\bf g}}

\title{On the Rare $ K \rightarrow \pi l^+ l^-$ Decays}

\author{David Greynat \address[CPT]{Centre  de Physique Th{\'e}orique~CNRS-Luminy, Case 907 F-13288 Marseille Cedex 9, France \\ CNRS UMR 6207} \thanks{Talk given at the $11^{th}$ High-Energy Physics International Conference on Quantum Chromodynamics, 5-10 July (2004), Montpellier (France).}}

\runtitle{On the rare $ K \rightarrow \pi \bar l l$ decays}
\runauthor{D. Greynat}

\begin{document}

\begin{abstract}
In this talk, we present our recent work ~\cite{FGEdeR04} on the $K\ra\pi \bar{l} l$ decays in a combined framework of chiral perturbation theory and  Large--$N_c$ QCD under the dominance of a minimal narrow resonance structure. The proposed description reproduces very well the experimental Br$(K^+\ra\pi^+ e^+ e^-)$ and Br$(K_S\ra\pi^0 e^+ e^-)$. Predictions for the $K\ra\pi~\mu^{+}\mu^{-}$ modes are also obtained and we can conclude to the constructive type for the interference between the {\it direct} and {\it indirect} CP--violation amplitudes in $K_L\ra\pi^0 e^+ e^-$. 
\vspace{1pc}
\end{abstract}
\maketitle

\section{Introduction}\lbl{int}

The analysis of $K\ra \pi\gamma^*\ra \pi l^{+}l^{-}$ decays within the framework of chiral perturbation theory ($\chi$PT) was first made in refs.~\cite{EPR87,EPR88}. To lowest non trivial order in the chiral expansion, the corresponding decay amplitudes get contributions both from chiral one loop graphs, and from tree level contributions of local operators coming from  the relevant effective Lagrangian $\Delta S =1$ at $\cO (p^4)$. In fact, in order to combine refs.~\cite{EPR87,EPR88} with our new theoretical view, it is more convenient to rewrite this Lagrangian as 
\begin{align} 
&\hspace{-1cm}\cL_{\rm eff}^{\Delta S=1}(x) \doteq \nonumber\\
&\hspace{-1cm}- G_8  \tr\left(\lambda L_{\mu}L^{\mu} \right) + eG_8F_{0}^2 A_{\mu}\tr[\lambda(L^{\mu}\Delta+\Delta L^{\mu})]\nonumber\\ 
&\hspace{-1cm}-  \frac{ie}{3F_0^2}G_8 F^{\mu\nu}({\bfw}_{1}-{\bfw}_{2})\; \tr \left( \lambda L_\mu L_\nu \right) \nonumber\\
&\hspace{-1cm}-  \frac{ie}{F_0^2}G_8 {\bfw}_{2}\, \tr(\lambda L_\mu \hat Q L_\nu) + \rm{h.c.}\lbl{efflnex}
\end{align}
where $U(x)$ is the matrix field which collects the Goldstone fields ($\pi$'s, $K$'s and $\eta$), $A_{\mu}$ is an external electromagnetic field source, $F^{\mu\nu}$ the  corresponding electromagnetic field strength tensor and 
\begin{flalign}
&\hspace{-0.8cm}L_{\mu}(x)=-iF_{0}^2 U^{\dagger}(x)\partial_{\mu}U(x)\;,\\
&\hspace{-0.8cm}\Delta(x)=U^{\dagger}(x)[\hat{Q},U(x)]\;,\; G_8=\scriptstyle{\frac{G_{F}}{\sqrt{2}}\,V_{\rm ud}^{\phantom{\ast}}V_{\rm us}^{\ast}\,\bfg_8}\;,\\
&\hspace{-0.8cm}\hat Q = \rm{diag}(1,0,0) \;,\; (\lambda)_{ij} = \delta_{3i}\delta_{2j}\;.
\end{flalign}
As can be show easily $\tilde {\bf w}$ is $\cO(N_c^0)$ and $\bfw_2$ is $\cO(N_c)$. Obviously our aim is to obtained values for this two coupling constants. Let us remind some basic facts concerning these decays. 


\section{$K\ra \pi l\bar{l}$ Decays to $O(p^4)$ in the Chiral Expansion}

In full generality, one can predict from   ref.~\cite{EPR87} the $K^{+}\ra \pi^{+} l^{+}l^{-}$ decay rates ($l=e,\mu$) as a function of the scale--invariant combination of coupling constants
\begin{align}
\hspace{-1cm}{\bfw}_{+}= &-\frac{(4\pi)^{2}}{3} \,[\tilde {\bf w}+3({\bfw}_{2}-4\mbox{\bf L}_{9})]\nonumber\\
&\hspace{2cm}-\frac{1}{6}\log\frac{M_{K}^2 m_{\pi}^{2}}{\nu^4}\label{eq:kppiplplmw}\,,
\end{align}

The predicted decay rate $\Gamma(K^+\ra\pi^+ e^+ e^-)$ is a quadratic function of ${\bfw}_{+}$, then there are two solutions to reproduce the experimental branching ratio~\cite{K+dacrate} (for a value of the overall constant ${\bf g}_8=3.3$):
\begin{equation}\lbl{K+rate}
{\rm Br}(K^+\ra\pi^+ e^+ e^-)=(2.88\pm0.13)\times 10^{-7}\,,	
\end{equation}
\begin{equation}
\lbl{solsK+}{\bfw}_{+}=1.69\pm 0.03 \;\; \rm{or}\;\; {\bfw}_{+}=-1.10\pm 0.03\;.
\end{equation}

The $K_{S}\ra \pi^{0}e^{+}e^{-}$ decay rate  brings in another scale--invariant combination of constants:
\be \lbl{eq:ws}
{\bfw}_{s}= -\frac{1}{3}(4\pi)^{2}\,\tilde {\bf w}-\frac{1}{3}\log\frac{M_{K}^{2}}{\nu^2}\,,
\ee
and it is also quadratic in ${\bfw}_{s}$. From the recent result on this mode, reported by the NA48 collaboration at CERN~\cite{NA48}:
\begin{align}
&\hspace{-0.8cm}{\rm Br}\left(K_S \rightarrow \pi^0 e^+ e^-\right)= \nonumber \\
&\left[5.8^{+2.8}_{-2.3} (\rm stat.) \pm 0.8 (\rm syst.) \right] \times 10^{-9}\lbl{K0rate}\,,
\end{align} 
one obtains the two solutions for ${\bfw}_{s}$
\begin{equation}\lbl{solsK0}
{\bfw}_{s}= 2.56 ^{+0.50}_{-0.53} \quad \rm{or} \quad 	{\bfw}_{s}= -1.90 ^{+0.53}_{-0.50}\,\,.
\end{equation}

At the same $\cO(p^4)$ in the chiral expansion, the branching ratio for the $K_L\ra \pi^0 e^+ e^-$ transition induced by CP--violation  reads as follows
\begin{align}
&\hspace{-1cm}{\rm Br}\left(K_L \rightarrow \pi^0 e^+ e^-\right)\vert_{\rm\tiny CPV}= \nonumber \\
&\hspace{-1cm}\left[(2.4\pm 0.2) \left(\frac{{\rm Im}\lambda_t}{10^{-4}}\right)^{2}+(3.9\pm 0.1)\,\left(\frac{1}{3}-{\bfw}_s\right)^2 \right. \nonumber\\
&\hspace{-0.5cm}\left.+ (3.1\pm 0.2)\, \frac{{\rm Im}\lambda_t}{10^{-4}}\left(\frac{1}{3}-{\bfw}_s\right)\right]\times 10^{-12}\lbl{KSCPV}\,.	
\end{align}

Here, the first term is the one induced by the {\it direct} source, the second one by the {\it indirect} source and the third one the {\it interference} term. With~\cite{Baetal03} $\Imm\lambda_t= (1.36\pm 0.12)\times 10^{-4}$, the interference is constructive for the negative solution in Eq.~\rf{solsK0}. 

The four solutions obtained in Eqs.~\rf{solsK+} and \rf{solsK0} define four different straight lines in the plane of the coupling constants ${\bfw}_{2}-4{\bf L}_9$ and ${\bf \tilde{w}}$, as illustrated in Fig.~1 below.  We next want to discuss which of these four solutions, if any, may be favored by theoretical arguments.

\begin{figure}[h]
\begin{center}
\includegraphics[width=0.47\textwidth]{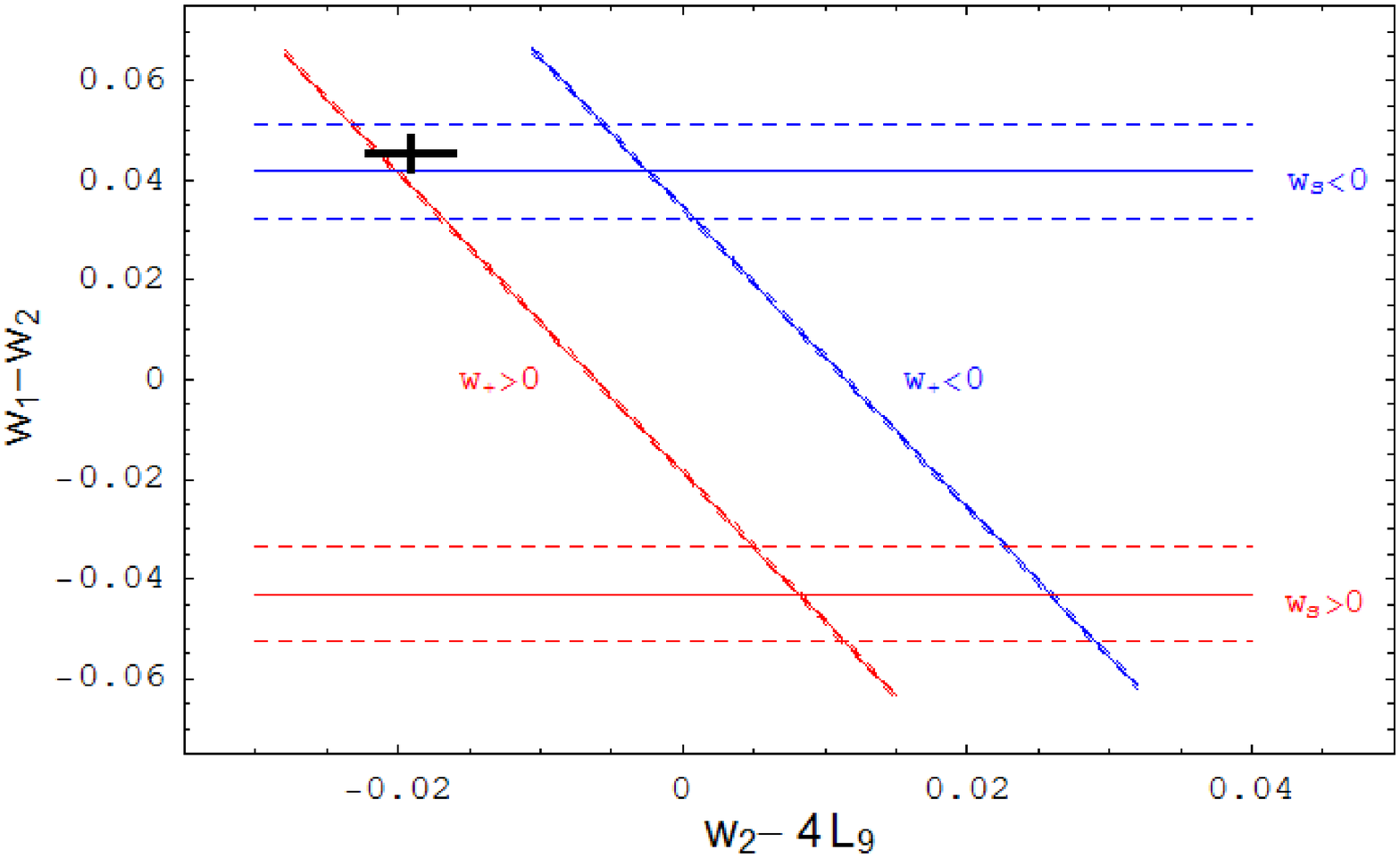}
\end{center}
{\bf Fig.~1} {\it The four possible values of the couplings at $\cO(p^4)$ in the chiral expansion are compatible with the experiments. The cross in this figure  corresponds to the values in Eqs.~\rf{wtildepred} and \rf{w24L9pred}. }
\end{figure}

\section{Theoretical Considerations}

\subsection{The Octet Dominance Hypothesis}

In ref.~\cite{EPR87}, it was suggested that  the couplings ${\bfw}_1$ and ${\bfw}_2$ may satisfy the same symmetry properties as the chiral logarithms generated by the one loop calculation. This selects the octet channel in the transition amplitudes as the only possible channel and leads to the relation 
\begin{equation}\lbl{odh}
{\bfw}_2=4{\bf L}_9\;{\mbox{\rm\footnotesize Octet Dominance Hypothesis (ODH)}}\,.
\end{equation}

We now want to show how this {\it hypothesis} can in fact be justified within a simple dynamical framework of resonance dominance, rooted in Large--$N_c$ QCD. For that, let us reduced the Lagrangian in Eq.~\rf{efflnex} at the minimum of one Goldstone field component:
\begin{align}
&\hspace{-0.8cm}\cL_{\rm eff}^{\Delta S=1}(x)  =\nonumber\\
&\hspace{-0.8cm} G_8ie \,{\bfw}_{2} \partial_{\nu}F^{\nu\mu}\tr[\lambda(\Phi\hat{Q}\partial_{\mu}\Phi-\partial_{\mu}\Phi\hat{Q}\Phi)] \;\scriptstyle{+ \cdots} \label{efflfield}
\end{align}
showing that the two--field content which in the term modulated by $ {\bfw}_{2}$ couples to $\partial_{\nu}F^{\nu\mu}$  is exactly the same as the one which couples to the gauge field $A^{\mu}$  in the lowest $\cO(p^2)$ Lagrangian and then, they are cancelled ~\cite{EPR87}. The cancellation is expected because of the mismatch between the minimum number of powers of external momenta required by gauge invariance and the powers of momenta that the lowest order effective chiral Lagrangian can provide. As we shall next explain, it is the reflect of the dynamics of this cancellation which, to a first approximation, is also at the origin of  the relation $\bfw_2=4\mbox{\bf L}_{9}$.
	
The hadronic electromagnetic interaction reads as follows
\be
\cL_{\rm em}(x)=-ie\left( A^{\mu}-\frac{2{\bf L}_{9}}{F_0^2} \partial_{\nu} F^{\nu\mu}\right) \tr(\hat{Q}\Phi\stackrel{\leftrightarrow}{\partial_{\mu}}\Phi) \;\scriptstyle{+\cdots}\,.
\ee
We can recognize here an electromagnetic form factor to the charged Goldstone bosons that begins as 
\be
F_{\rm em}(Q^2)=1-\frac{2{\bf L}_{9}}{F_{0}^2}Q^2 + \cdots 
\ee 
In the {\it minimal hadronic approximation} (MHA) to Large--$N_c$ QCD, the form factor in question is saturated by the lowest order pole
i.e. the $\rho(770)$~:
\be
\label{Fem}
 F_{\rm em}(Q^2)=\frac{M_{\rho}^2}{M_{\rho}^2+Q^2}\;\Rightarrow\;{\bf L}_{9}=\frac{F_{0}^2}{2M_{\rho}^2}\,.
\ee

It is well known that this reproduces the observed slope rather well. By the same argument in Eq.~(\ref{efflfield}), we have an electroweak form factor
\be
F_{\rm ew}(Q^2)= 1-\frac{{\bf w}_{2}}{2F_{0}^2}Q^2 + \cdots \;.
\ee

Here, however, the underlying $\Delta S=1$ form factor structure can have contributions both from the $\rho$ and the $K^*(892)$~: 
\begin{equation}
F_{\rm ew}(Q^2)= \frac{\alpha M_{\rho}^2}{M_{\rho}^2+Q^2} + \frac{\beta M_{K^*}^2}{M_{K^*}^2+Q^2},
\end{equation}
with $\alpha+\beta=1$ because at $Q^2\ra 0$ the form factor is normalized to one by gauge invariance. This fixes the slope to 
\begin{equation}
\label{w2surF}
 \frac{{\bf w}_2}{2F_0^2} =\left(\frac{\alpha}{M_{\rho}^2}+\frac{ \beta}{M_{K^*}^2}\right)\,.
\end{equation}
If, furthermore, one assumes the chiral limit where $M_\rho =M_{K^*}$, there follows then combining (\ref{Fem}) and (\ref{w2surF}), the ODH relation in Eq.~\rf{odh};
a result which, as can be seen in Fig.~1, favours the solution where both ${\bf w}_{+}$ and ${\bf w}_{s}$ are negative, and the interference term in Eq.~\rf{KSCPV} is then constructive.

\subsection{Beyond the $\cO(p^4)$ in $\chi$PT}

Here, we want to show that it is possible to understand the observed $K^+ \rightarrow \pi^+ l^+l^-$ spectrum within a simple MHA picture of Large--$N_c$ QCD which goes beyond the $\cO(p^4)$ framework of $\chi$PT but, contrary to the proposals in refs.~\cite{DEIP98,BDI03}, it does not enlarge the number of free parameters.

\begin{figure}[h]
\begin{center}
\includegraphics[width=0.46\textwidth]{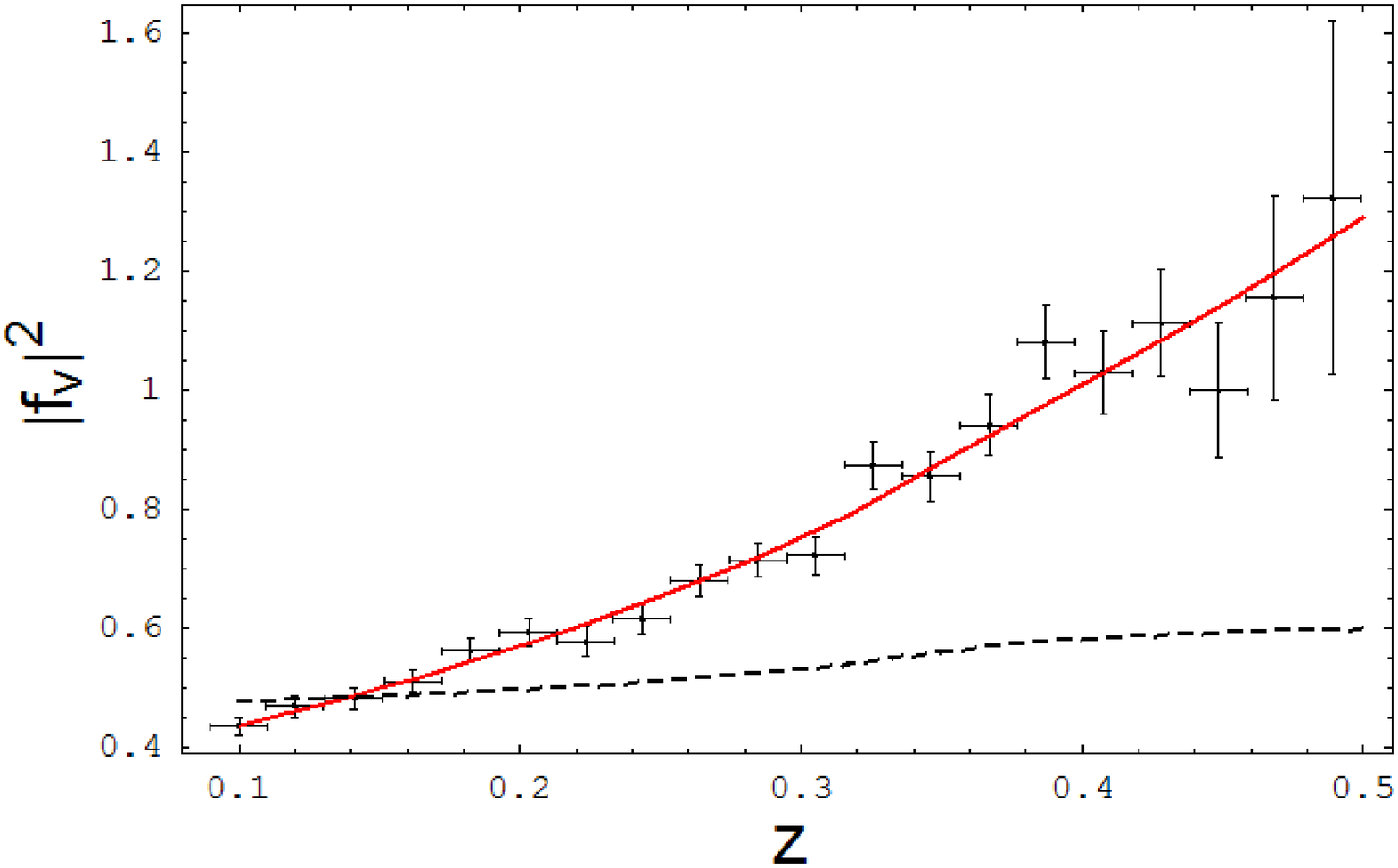}
\end{center}
{\bf Fig.~2} {\it Plot of the form factor $\left|f_V(z)\right|^2$ versus the invariant mass squared of the $e^+ e^-$ pair normalized to $M_{K}^2$. The crosses are the experimental points ~\cite{zeller}; the dotted curve is the (best) leading $\cO(p^4)$ prediction ($\bfw_+ >$0); the continuous line is the fit of the form factor in Eq.~\rf{fff} below.}
\end{figure}

Following the ideas  developed in the previous subsection, we propose a very simple generalization of the $\cO(p^4)$ form factor ~\cite{EPR87}:
\begin{equation}
f_V(z)  = \frac{G_8}{G_F} \left\{ \frac{1}{3}-w_+ - \frac{1}{60}z - \chi(z) \right\}\;.
\end{equation}

We keep the lowest order chiral loop contribution as the leading manifestation of the Goldstone dynamics, but replace the local couplings $\bfw_2 -4{\bf L}_9$ and ${\tilde{\bf w}}$ in ${\bf w}_{+}$ by the minimal resonance structure. The form factor we propose is ~\cite{FGEdeR04}
\begin{align}
\lbl{fff}
&\hspace{-0.8cm}f_V(z) = \frac{G_8}{G_F} \left\{  \frac{(4\pi)^2}{3} \left[{\tilde{\bf w}}\frac{M_{\rho}^2}{M_{\rho}^2-M_{K}^2 z} \right. \right. \nonumber\\ 
&\hspace{-0.4cm}\left.\left.+ 6   F_{\pi}^2 \beta \frac{M_\rho^2 - M_{K^*}^2}{\left(M_\rho^2 - M_K^2 z\right)\left(M_{K^*}^2-M_K^2 z\right)}\right] \right. \nonumber\\
&\hspace{-0.8cm}\left. + \frac{1}{6} \ln \left(\frac{M_K^2 m_\pi^2}{M_\rho^4}\right) +\frac{1}{3} - \frac{1}{60}z - \chi(z) \right\}\,,
\end{align}
where $\chi(z)=\phi_{\pi}(z)-\phi_{\pi}(0)$. 

With $\tilde {\bf w}$ and $\beta$ left as free parameters, we make a least squared fit to the experimental points in Fig.~2. The result is the continuous curve shown in the same figure, which corresponds to a $\chi_{\mbox{\rm\tiny min.}}^2=13.0$ for 18 degrees of freedom. The fitted values (using ${\bf g}_8=3.3$ and $F_{\pi}=92.4~\MeV$) are 
\begin{equation}\lbl{wtildepred}
	\tilde {\bf w}=0.045\pm 0.003\qquad\annd\qquad \beta=2.8\pm 0.1\,;
\end{equation}
and therefore
\begin{equation}\lbl{w24L9pred}
		\bfw_2 -4{\bf L}_9=-0.019\pm 0.003\,.
\end{equation}
These are the values which correspond to the cross in Fig.~1 above. The fitted value for $\tilde{\bf w}$ results in a negative value for $\bfw_s$ in Eq.~\rf{eq:ws} 
\begin{equation}
\lbl{fitws}\bfw_s=-2.1\pm 0.2\,,
\end{equation}
 which corresponds to the branching ratios (for experimental values see \cite{K+dacrate}\cite{Moriond})
\begin{align}
&\hspace{-0.8cm}{\rm Br}\left(K_S \rightarrow \pi^0 e^+ e^-\right)  =   (7.7\pm 1.0)\times 10^{-9}\,, \\
&\hspace{-0.8cm}{\rm Br}\left(K_S \rightarrow \pi^0 e^+ e^-\right)\vert_{\scriptscriptstyle  >165 {\tiny \MeV}}  =  (4.3\pm 0.6)\times 10^{-9}
\end{align}
and with \rf{w24L9pred} to 
\begin{align}
&\hspace{-0.8cm}{\rm Br}\left(K^+ \rightarrow \pi^+ \mu^+ \mu^-\right) =   (1.7\pm 0.2)\times 10^{-9}\,.	
\end{align}

Finally, the resulting negative value for ${\bf w}_s$ in Eq.~\rf{fitws}, implies a constructive interference in Eq.~\rf{KSCPV} with a predicted branching ratio
\begin{equation}\lbl{KLCPVP}
{\rm Br}\left(K_L \rightarrow \pi^0 e^+ e^-\right)\vert_{\rm\tiny CPV}=(3.7\pm 0.4)\times 10^{-11}\,, 
\end{equation}
where we have used~\cite{Baetal03} $\Imm\lambda_t= (1.36\pm 0.12)\times 10^{-4}$.

\section{Conclusions}

Earlier analyses of $K\ra\pi~e^+ e^-$ decays within the framework of $\chi$PT have been extended beyond the predictions of $\cO(p^4)$, by replacing the local couplings which appear at that order by their underlying  narrow resonance structure in the spirit of the MHA to Large-$N_c$ QCD. The resulting modification of the $\cO(p^4)$ form factor is very simple and does not add new free parameters. It reproduces very well both the experimental decay rate and the invariant $e^+ e^-$ mass spectrum. The predicted Br$(K_S\ra\pi^0 e^+ e^-)$ and  Br$(K_S\ra\pi^0 \mu^+ \mu^-)$ are, within errors, consistent with the recently reported result from the NA48 collaboration. The predicted interference between the {\it direct} and {\it indirect} CP--violation amplitudes in $K_L\ra\pi^0 e^+ e^-$ is constructive, with an expected branching ratio (see Eq.~\rf{KLCPVP}) within reach of a dedicated experiment.

\end{document}